\documentclass[twocolumn,
showpacs,preprintnumbers,superscriptaddress,amsmath,amssymb,floatfix,aps]{revtex4}

\usepackage[dvips]{graphicx}% Include figure files
\usepackage{dcolumn}% Align table columns on decimal point
\usepackage{bm}% bold math
\begin{document}

%\preprint{APS/123-QED}

\title{Casimir-Lifshitz force out of thermal equilibrium and asymptotic non-additivity}

\author{Mauro Antezza}
%\email[Corresponding author: ]{antezza@science.unitn.it}
\affiliation{Dipartimento di Fisica, Universit\`a di Trento
and CNR-INFM R\&D Center on Bose-Einstein Condensation, Via Sommarive 14, I-38050 Povo, Trento, Italy}
\author{Lev P. Pitaevskii}
\affiliation{Dipartimento di Fisica, Universit\`a di Trento
and CNR-INFM R\&D Center on Bose-Einstein Condensation, Via Sommarive 14, I-38050 Povo, Trento, Italy}
\affiliation{Kapitza Institute for Physical Problems, ul. Kosygina 2, 119334 Moscow,
Russia}
\author{Sandro
Stringari}
\affiliation{Dipartimento di Fisica, Universit\`a di Trento
and CNR-INFM R\&D Center on Bose-Einstein Condensation, Via Sommarive 14, I-38050 Povo, Trento, Italy}
\author{Vitaly B. Svetovoy}
\affiliation{MESA+ Research Institute, University of Twente, PO 217, 7500 AE Enschede, The Netherlands}

\date{\today}

\begin{abstract}
We investigate the force acting between two parallel plates held at different temperatures. The force reproduces, as limiting cases, the well known Casimir-Lifshitz surface-surface force at thermal equilibrium and the surface-atom force out of thermal equilibrium recently derived by M. Antezza \emph{et al.}, Phys. Rev. Lett. {\bf 95}, 113202 (2005). 
The  asymptotic behavior of the force at large distances is  explicitly discussed. In particular when one of the two bodies is a rarefied gas the force is not additive, being proportional to the square root of the density. Nontrivial cross-over regions at large distances are also identified.
\end{abstract}

\maketitle

The study of the thermal fluctuations of the electromagnetic field and of their effects on the force acting on surfaces and atoms is a longstanding subject of theoretical research starting from the seminal Lifshitz paper \cite{Lifshitz_56} (see also \cite{LP}). The elusive nature of the thermal component of the force  follows from the fact that thermal effects becomes visible only at large distances, of the order of  the photon thermal wave length $\lambda_T=\hbar c/k_B T$, where they prevail on the Casimir force originating from the $T=0$ quantum fluctuations of the field. At room temperature the thermal wave length corresponds to about $7$ microns, a distance at which   both the Casimir and the thermal forces are very weak and difficult to reveal experimentally. The existence of thermal effects has been experimentally demonstrated only recently by the JILA experiment  \cite{marchmeeting}, by measuring the frequency shift of the center of mass motion of  an ultracold atomic cloud located  at a distance of a few microns from a dielectric substrate \cite{articolo1,eric05}.

Thermal fluctuations determine the asymptotic, large distance behavior of the electromagnetic pressure which takes the
Lifshitz form \cite{LP}
\begin{equation}
P_{\textrm{th}}^{\textrm{eq}}(T,l)=\frac{k_BT}{16\pi
l^3}\!\int_0^{\infty}\!\!\!x^2\!\left[\frac{(\varepsilon_{10}+1)(\varepsilon_{20}+1)}
{(\varepsilon_{10}-1)(\varepsilon_{20}-1)}\;e^x-1\right]^{-1}\!\!\textrm{d}x,
\label{LifLDbb}
\end{equation}
in the case of two parallel surfaces separated by a distance $l$.
Here $\varepsilon_{10}=\varepsilon_{1}(0)$ and $\varepsilon_{20}=\varepsilon_{2}(0)$  are the static dielectric constants of the two materials  and  $T$ is the temperature of the system.
When the system is not in thermal equilibrium  the pressure is expected to exhibit a  different behavior. In particular in recent papers the Trento team \cite{articolo2,AntezzaJPA,PitaevskiiJPA} has shown that the surface-atom force out of thermal equilibrium exhibits a new asymptotic behavior at large distances.
With respect to that at equilibrium, the new force  exhibits a slower dependence on the distance and a stronger dependence on the temperature, making its experimental detection more accessible as demonstrated in the experiment of \cite{marchmeeting}.

The purpose of the present work is to investigate the 
behavior of the force out of thermal equilibrium in the case of two
parallel surfaces. The general goal is to better understand the role of thermal fluctuations which is not fully exploited at thermal equilibrium, being masked by peculiar cancellation effects between propagating and evanescent waves  \cite{articolo2,PitaevskiiJPA}. In particular we  address the following
questions: a) how is the Lifshitz  law - and its asymptotic limit
(\ref{LifLDbb}) -  modified if the temperatures of the two
bodies are different? b) Can one recover the results of
\cite{articolo2} for the surface-atom force when one 
body is made of a very dilute material corresponding to a
gaseous phase with $(\varepsilon-1)\rightarrow0$ ? 

Let us consider two parallel dielectric half spaces \emph{locally}
at  thermal equilibrium with different temperatures and separated by
a distance $l$. In our configuration the left-side (right-side) body
has a complex dielectric function $\varepsilon_1(\omega)$
($\varepsilon_2(\omega)$) and is  held  at
temperature $T_1$ ($T_2$), the whole system being in a
\emph{stationary configuration}. We assume that each body fills an
infinite half-space (see, however, discussion after Eq.(\ref{crossingeq})). In practice this means that the bodies are thick compared to the penetration depth of the thermal radiation. In such conditions the presence of the remote surfaces of the bodies results only in a $l$-independent contribution to pressure, which will not be considered in this Letter. 

Let us assume that
the separation between the bodies is in the $z-$direction. Then the
electromagnetic pressure between them is given by the average 
$P^{\textrm{neq}}(T_1,T_2,l)=\langle T_{zz}({\bf r},t) \rangle$, 
where $T_{zz}({\bf r},t)
=\left(E_z^2-E_x^2-E_y^2+B_z^2-B_x^2-B_y^2\right)/8\pi$ is the $zz$
component of the Maxwell stress tensor \cite{LLPCM} in the vacuum
gap. In this work we focus on the thermal component $P_{\textrm{th}}$ of the force,
defined by $P(T_1,T_2,l)=P_{0}(l)+P_{\textrm{th}}(T_1,T_2,l)$, where
$P_0$ is the $T=0$ quantum pressure originating from the vacuum fluctuations of the field \cite{LP}.

The  electromagnetic field in the vacuum gap physically originates \cite{LLPCM}
from   the  fluctuating polarization fields ${\bf
P}[\omega;{\bf r}]$ in the two half-spaces whose correlations inside
each body are described by the fluctuation-dissipation theorem
\begin{eqnarray}
\lefteqn{\!\!\!\!\!\left\langle
P_k[\omega,{\textbf{r}}]P_l^*[\omega',{\textbf{r}}']\right\rangle=}\notag\\
&&\!\!\!\!\frac{\hbar\;\varepsilon''_{1,2}(\omega)}{2}\coth\left(\frac{\hbar\omega}{2k_B
T_{1,2}}\right)
\delta(\omega-\omega')\delta({\bf r}-{\bf r}')\delta_{kl},
\label{FDTPol}
\end{eqnarray}
where $\varepsilon_{1,2}''(\omega)$ is the imaginary part of the
dielectric function of  the two materials.
%===================================================================================================
\begin{figure}[ptb]
\begin{center}
\includegraphics[width=0.45\textwidth]{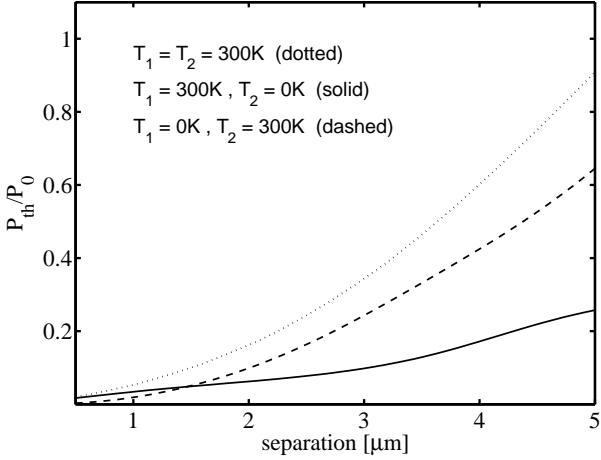}
\vspace{-0.3cm}
\caption{Relative contribution of the thermal component with respect to the zero-temperature component of the pressure  between two different materials: Fused Silica (SiO$_2$, body 1) and Silicon (Si, body 2).} 
\label{Fig:fig1}
\vspace{-0.5cm}
\end{center}
\end{figure}
%=================================================================================================== 
Due to the presence of the $\delta({\bf r}-{\bf r}')$ factor  these correlations are  \emph{local} \cite{Spatial} so that the effects
of the fluctuations originating from the two half-spaces  add
incoherently. Assumption (\ref{FDTPol}) (local source
hypothesis) was first used in \cite{Polder} and  represents the  starting point of our analysis allowing
for an explicit calculation of the electromagnetic field also if the
system is not in global thermal equilibrium
\cite{Henkel1,articolo2,AntezzaJPA,PitaevskiiJPA,Greffet05}. The
electric field at the point ${\bf r}$ in the gap can be in fact
expressed in terms of the source polarization field via the
convolution ${\bf E}\left[\omega;{\bf r}\right]=\int {\bf
{G}}\left[\omega;{\bf r},{\bf r}'\right]\;\cdot {\bf
P}\left[\omega;{\bf r}'\right]\;\textrm{d}{\bf r}'$ of the Green
tensor \cite{SipeForm}, where the integration is performed over
the volume containing the sources at ${\bf r}'$.  At the same time
the  magnetic field is easily evaluated  using the Maxwell equation
${\bf B}[\omega;{\bf r}]=-i\nabla\wedge{\bf E}[\omega;{\bf r}]/k$, where $k=\omega/c$.
Then Eq. (\ref{FDTPol})  allows us to write 
the thermal pressure acting between the bodies as the sum of
two terms:
\begin{equation}
P_{\textrm{th}}^{\textrm{neq}}(T_1,T_2,l)= P_{\textrm{th}}^{\textrm{neq}}(T_1,0,l)+P_{\textrm{th}}^{\textrm{neq}}(0,T_2,l),
\label{ForceNeq}
\end{equation}
each of them corresponding to a configuration where only one of the
two materials is at  non-zero temperature. It is convenient to write 
\begin{eqnarray}
P_{\textrm{th}}^{\textrm{neq}}(T,0,l)&=&P_{\textrm{th}}^{\textrm{eq}}(
T,l)/2+\Delta P_{\textrm{th}}(T,l),
\label{neqPWEW1}\\
P_{\textrm{th}}^{\textrm{neq}}(0,T,l)&=&P_{\textrm{th}}^{\textrm{eq}}(
T,l)/2-\Delta P_{\textrm{th}}( T,l),
\label{neqPWEW2}
\end{eqnarray}
where $P_{\textrm{th}}^{\textrm{eq}}( T,l)$ is the Lifshitz pressure at
equilibrium \cite{LP}. If we write the electromagnetic wave vector in the vacuum gap as
${\bf k}=(Q_x,Q_y,q_z)$, whose longitudinal part has
modulus $Q=\sqrt{Q_x^2+Q_y^2}$, and $q_z=\sqrt{k^2-Q^2}$, it is possible to express the quantity $\Delta P_{\textrm{th}}$ as the sum 
$\Delta P_{\textrm{th}}=\Delta P_{\textrm{th}}^{\textrm{PW}}+\Delta
P_{\textrm{th}}^{\textrm{EW}}$, with
\begin{widetext}
\begin{eqnarray}
\Delta P_{\textrm{th}}^{\textrm{PW}}(T,l)&=&-\frac{\hbar}{4\pi^2}\int_0^{\infty}\textrm{d}\omega\frac{1}{e^{\hbar\omega/k_BT}-1}\int_0^k\textrm{d}Q\;Q\;q_z\;\sum_{\mu=s,p}\left(|r^{\mu}_{2}|^2-|r^{\mu}_{1}|^2\right)\;\left(\frac{1}{|D_{\mu}|^2}-\frac{1}{1-|r^{\mu}_{1}\;r^{\mu}_{2}|^2}\right),
\label{PWA}\\
\Delta P_{\textrm{th}}^{\textrm{EW}}(T,l)&=&\frac{\hbar}{2\pi^2}\int_0^{\infty}\textrm{d}\omega\frac{1}{e^{\hbar\omega/k_BT}-1}\int_k^{\infty}\textrm{d}Q\;Q\; \textrm{Im}q_z\;e^{-2l\textrm{Im}q_z}\sum_{\mu=s,p}\frac{\textrm{Im}\left(r^{\mu}_{1}\right)\textrm{Re}\left(r^{\mu}_{2}\right)-\textrm{Im}\left(r^{\mu}_{2}\right)\textrm{Re}\left(r^{\mu}_{1}\right)}{|D_{\mu}|^2},
\label{EWI}
\end{eqnarray}
\end{widetext}
where we have separated the effect of propagating waves (PW) from
that of  evanescent  waves (EW) corresponding, respectively, to real
and imaginary values of $q_z$. We have also  subtracted the $l$-independent contributions so that both Eqs.(\ref{PWA}) and (\ref{EWI}) vanish as $l\rightarrow\infty$. 
In Eqs. (\ref{PWA}) and (\ref{EWI}) the dielectric properties of the two materials enters through the
reflection Fresnel coefficients for the vacuum-dielectric interfaces
\begin{equation}
r^s_{m}=\frac{q_z-q_z^{(m)}}{q_z+q_z^{(m)}}\;\;,\;\;r^p_{m}=\frac{q_z\varepsilon_m-q_z^{(m)}}{q_z\varepsilon_m+q_z^{(m)}},
\label{rsrp}
\end{equation}
where $s$ and $p$ correspond to the transverse electric and magnetic
polarizations and
 $q_z^{(m)}=\sqrt{\varepsilon_{m}\;k^2-Q^2}$ is the $z$-th component of the wave
vector in the material $m=1,2$. The effect of multiple
reflections between bodies is accounted for by the denominator $D_{\mu}
=1-r^{\mu}_{1}r^{\mu}_{2}e^{2iq_zl}$. 
It is evident that the pressure  $P_{\textrm{th}}^{\textrm{neq}}(0,T,l)$ of
Eq.~(\ref{neqPWEW2}) can be obtained  from Eq.~(\ref{neqPWEW1}) by replacing  $r^{\mu}_{1}\leftrightarrow r^{\mu}_{2}$. 
It is also immediate to see that Eq.~(\ref{ForceNeq}) at thermal equilibrium $T_1=T_2\equiv T$ 
 provides  the well known equilibrium  pressure $P_{\textrm{th}}^{\textrm{eq}}( T,l)$. For 
identical bodies $r^{\mu}_{1}=r^{\mu}_{2}$, yielding $\Delta
P_{\textrm{th}}=0$, the pressure (\ref{ForceNeq}) is given by the expression
\begin{equation}
P_{\textrm{th}}^{\textrm{neq}}(T_1,T_2,l)=P_{\textrm{th}}^{\textrm{eq}}( T_1,l)/2+P_{\textrm{th}}^{\textrm{eq}}(T_2,l)/2.
\label{Dorof.}
\end{equation}
Equation (\ref{Dorof.}), previously obtained in \cite{Dorofeyev1},
is not however valid if the two materials are different. This is in
disagreement with the results of  \cite{Dorofeyev2}, where
(\ref{Dorof.}) was found to be valid in general \cite{Dorofeyevcomment}. In particular Eqs. (\ref{neqPWEW1}) and (\ref{neqPWEW2})   show that interchanging the temperatures of the two materials implies a change in the  force if the two materials are different. This clearly emerges from Fig.\ref{Fig:fig1}, where the results of a full calculation of the pressure between Fused Silica (body 1) and Silicon (body 2) are presented \cite{dielfunc}. It is interesting to note that the relative values of the pressure for the two non-equilibrium configurations (dashed and solid lines)  strongly depends on the temperatures of the two bodies, on the separation $l$, and on the positions of the resonances of the two dielectric functions $\varepsilon_{1,2}(\omega)$. One can also see that both values are smaller than the one at thermal equilibrium (dotted line). This it is not however always the case. In fact if one of the two bodies is rarefied the non equilibrium pressure can become larger than at equilibrium (see \cite{articolo2} and discussion below). 

In the following we will focus on the  behavior of the force at
large distances.  For this study we will consider only 
dielectric bodies (where the limit $\varepsilon_{1,2}(\omega\rightarrow 0)$ is finite), while the case of metallic bodies will be presented elsewhere.
The expansion  of Eq.(\ref{neqPWEW1}) results in the large distance behaviors
\begin{widetext}
\begin{eqnarray}
P_{\textrm{th}}^{\textrm{neq,PW}}(T,0,l)&=&\frac{k_BT}{l^3}\;\frac{\zeta(3)}{16\pi}\left[2-\frac{\sqrt{\varepsilon_{10}-1}-\sqrt{\varepsilon_{20}-1}}{\sqrt{\varepsilon_{10}-1}+\sqrt{\varepsilon_{20}-1}}-\frac{\varepsilon_{20}\sqrt{\varepsilon_{10}-1}-\varepsilon_{10}\sqrt{\varepsilon_{20}-1}}{\varepsilon_{20}\sqrt{\varepsilon_{10}-1}+\varepsilon_{10}\sqrt{\varepsilon_{20}-1}}\right],
\label{AsimPW}\\
P_{\textrm{th}}^{\textrm{neq,EW}}(T,0,l)&=&\frac{k_BT}{l^3}\;\frac{1}{8\pi^2}\int_0^{\infty}\textrm{d}t\int_0^{\infty}\textrm{d}x\;\frac{x^2\;e^{-x}}{t}\;\sum_{\mu=s,p}\;\frac{\textrm{Im}\left[r^{\mu}_{1}(t)\right]\textrm{Re}\left[r^{\mu}_{2}(t)\right]}{|1-r^{\mu}_{1}(t)r^{\mu}_{2}(t)\;e^{-x}|^2},
\label{AsimEW}
\end{eqnarray}
\end{widetext}
holding for $l\gg\lambda_T$ (if one of the bodies is rarefied the condition becomes more stringent, see Eq.(\ref{largel}) below). Here $\zeta(3)$ is the Riemann function,
 $r^{\mu}_{m}(t)$ are the Fresnel reflection coefficients calculated  from Eq.(\ref{rsrp}) setting the static approximation $\varepsilon_{m}=\varepsilon_{m0}$, and $Q^2=k^2\left(1+t^2\right)$. The PW and EW contributions of Eqs.  (\ref{AsimPW}) and (\ref{AsimEW})  are of the same order but have opposite signs, the PW term being attractive and the EW repulsive. This feature is not however general and at shorter distance the signs of the PW and EW terms can change as discussed below. At distances of the order of the thermal wavelength Eqs.(\ref{AsimPW}) and (\ref{AsimEW}), as well as the Lifshitz result (\ref{LifLDbb}), provide only a crude estimate of the pressure. For example
in the case of the Fused Silica ($\varepsilon_{10}\approx3.8$) - Silicon ($\varepsilon_{20}\approx11.7$)  configuration   at $l=5\mu$m, the asymptotic laws  overestimate the full calculation shown in Fig.(\ref{Fig:fig1})  by a factor $1.5$, $2$ and $3$ for the dashed, dotted and solid lines, respectively.

Since in this work we are also interested in recovering the
surface-atom force \cite{articolo2}  which is relevant for the recent experiments with ultracold gases \cite{marchmeeting}, it is useful to study the case in which the body $2$ is  cold and very rarefied, corresponding to small values of $(\varepsilon_2-1)=4\pi n \alpha$. Here $n$ is the density of the material $2$ and $\alpha$ is the dipole polarizability of the atoms.
The expansion of  pressure in Eq. (\ref{neqPWEW1}) should be performed through two limiting procedures: the large distance $l\rightarrow\infty$ and the diluteness  $(\varepsilon_2-1)\rightarrow 0$ conditions. It is crucial to identify the proper order of the two limits. 

One relevant asymptotic behavior  is obtained by {\it first} taking 
 the limit of large $l$ at fixed $\varepsilon_{2}$ (this yields, by the way, expressions (\ref{AsimPW}) and (\ref{AsimEW})), and {\it then} carrying out  the
limit of rarefied body. One finds the expression
\begin{equation}
P_{\textrm{th}}^{\text{neq}}(T,0,l)=\frac{k_BT\;C}{l^3}\frac{\varepsilon _{10}+1}{\sqrt{\varepsilon
_{10}-1}}\sqrt{\varepsilon _{20}-1} ,
\label{dFfisncca}
\end{equation}
for the total pressure,
where $C=C_{\textrm{PW}}+C_{\textrm{EW}}\approx3.83~\cdot~10^{-2}$, is a numerical factor with $C_{\textrm{PW}}=\zeta(3)/8\pi\approx 4.78~\cdot~10^{-2}$ and $C_{\textrm{EW}}\approx -0.96~\cdot~10^{-2}$.
 
The peculiar $\sqrt{\varepsilon _{20}-1}$ dependence of (\ref{dFfisncca}) means that
 the pressure acting on the atoms of the substrate $2$ \emph{is not additive}. Additivity
would in fact require a linear dependence on $(\varepsilon _{20}-1)$, and hence on the gas density $n$, as happens for the Lifshitz pressure (\ref{LifLDbb})  as $(\varepsilon _{20}-1)\to 0$:
\begin{equation}
P_{\textrm{th}}^{\textrm{eq}}(T,l) = \frac{k_BT}{16\pi
l^3}\frac{\varepsilon_{10}-1}{\varepsilon_{10}+1}\;(\varepsilon_{20}-1).
\label{LifLDbbab}
\end{equation}
The non additivity of
the pressure (\ref{dFfisncca}) follows from the fact that for large $l$
 the main contribution in the force is produced by the \emph{%
grazing waves} incident on the interface of the material $2$ from the vacuum gap at small angles corresponding to small values of $q_z/k\leq \sqrt{\varepsilon _{20}-1}$. Hence
the  reflection coefficients from the body $2$ is not small even at
small $\varepsilon _{20}-1$ and the body cannot be considered
dilute from the electrodynamic point of view. This is a peculiarity of the
non-equilibrium situation. In fact at equilibrium this anomalous contribution is canceled
 by the waves impinging the interface  from the interior of the dielectric $2$,
 close to the angle of total reflection. 
Notice that result (\ref{dFfisncca}) is valid at the condition
\begin{equation}
l\gg \lambda_T/\sqrt{\varepsilon_{20}-1},
\label{largel}
\end{equation}
which  becomes stronger and stronger as $(\varepsilon _{20}-1)~\to~0$.

The second limiting procedure is  obtained by {\it first} taking the expansion of (\ref{neqPWEW1}) for small values of 
$(\varepsilon_2-1)$, and {\it then} carrying out  the
limit of large distances. In this case the relevant wavevectors satisfy the condition  $ q_z/k \gg \sqrt{\varepsilon _{20}-1}$, and the PW component produces a contribution identical to the EW one, yielding
\begin{equation}
P_{\textrm{th}}^{\text{neq}}(T,0,l) =\frac{(k_BT)^2}{24\;l^{2}\;
c\hbar}\frac{\varepsilon _{10}+1}{\sqrt{\varepsilon _{10}-1}}\; (\varepsilon_{20}-1).
\label{art2}
\end{equation}
Result (\ref{art2}) holds in the distance range complementary to (\ref{largel})
\begin{equation}
\lambda_T \ll l \ll \lambda_T/\sqrt{\varepsilon _{20}-1}.
\label{smalle2}
\end{equation}
In deriving Eq.(\ref{art2}) we also replaced $\varepsilon _{1,2}\left(
\omega \right) $ with their static values $\varepsilon _{m0}$. This is justified if $k_BT$ is much smaller than the lowest resonances of both the body $1$ and the atoms of the dilute body $2$. It is worth noting that the interval (\ref{smalle2}) practically disappears for dense dielectrics. 
%===================================================================================================
\begin{figure}[ptb]
\begin{center}
\includegraphics[width=0.38\textwidth]{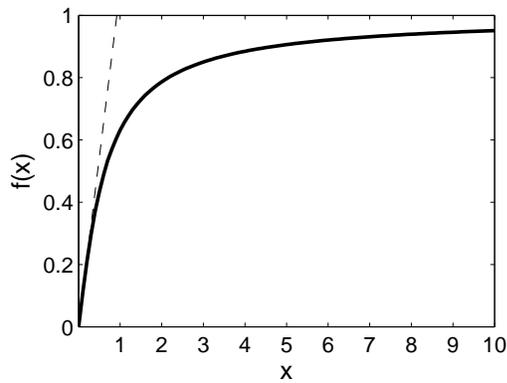}
\vspace{-0.3cm}
\caption{Dimensionless function $f(x)$ (see Eq.(\ref{crossingeq})).
The dashed line gives the asymptotic limit at small $x$.} 
\label{Fig:fig2}
\vspace{-0.7cm}
\end{center}
\end{figure}
%=================================================================================================== 

It is also worth noticing  that, due to the diluteness condition $(\varepsilon_{20}-1)\ll1$ and as a consequence of Eqs.(\ref{neqPWEW1}) - (\ref{neqPWEW2}), in both regions (\ref{largel}) and (\ref{smalle2}) the term $\Delta P_{\textrm{th}}$ gives the leading contribution to the pressure and consequently the $l-$dependent  interaction between the two bodies will be attractive if $T_1>T_2$ and repulsive in the opposite case.

The transition between the two regimes (\ref{dFfisncca}) and (\ref{art2}) can be investigated performing the diluteness limit  $(\varepsilon_{20}-1)\rightarrow 0$ in Eq.(\ref{neqPWEW1}), for a fixed value of the dimensionless variable $x=l\sqrt{\varepsilon_{20}-1}/\lambda_T$. The results are reported in Fig.\ref{Fig:fig2}, where the thermal pressure $P_{\text{th}}^{\text{neq}}(T,0,l)$ is plotted in units of the asymptotic behavior (\ref{dFfisncca}):
\begin{equation}
P_{\text{th}}^{\text{neq}}(T,0,l)= \frac{k_BT\;C}{l^3}\frac{\varepsilon _{10}+1}{\sqrt{\varepsilon
_{10}-1}}\sqrt{\varepsilon _{20}-1}  \;f(x).
\label{crossingeq}
\end{equation}
Here $f(x)$ is a dimensionless function of the variable $x$. When $x\rightarrow\infty$ (regime (\ref{dFfisncca})) one has $f(x)\rightarrow 1$, while when $x\rightarrow0$ (regime (\ref{art2})) one find $f(x)\rightarrow x/24C\approx1.09x$.

In order to recover the asymptotic result of \cite{articolo2} for the surface-atom force out of thermal equilibrium  it is crucial to follow the second limiting procedure, leading to result (\ref{art2}). In this case, however, the PW term
must be omitted since the atomic gas occupies a finite region of space and does not absorb the thermal radiation. Using the formalism of the present work this corresponds to treating the body $2$ as a slab of rarefied gas of thickness $L$ for which   
one should also take into account the force
acting on its remote surface. 
In the absence of absorption \cite{absorption} it is possible to show that, including refraction at the remote surface, 
the PW pressure becomes vanishingly small, of order  $\left( \varepsilon
_{20}-1\right) ^{2}$ with respect to Eq. (\ref{art2}). 
In this case the EW contribution, which is $1/2$ of  (\ref{art2}),  provides the total pressure acting on the gas and is exactly  equivalent to  equation (12) of \cite{articolo2} for the surface-atom force. Notice that in the derivation of  \cite{articolo2} the leading role of the EW term was stressed from the very beginning.

In conclusion we have generalized the Casimir-Lifshitz surface-surface force to configurations out of thermal equilibrium and calculated the corresponding asymptotic behavior. When one of the two bodies is a rarefied gas  a cross-over region emerges where the pressure changes from a $T^2/l^2$ behavior, characterizing the surface-atom interaction, to a region at very large distances where the pressure behaves like $T/l^3$ and is no longer additive.

\emph{Acknowledgments}: we gratefully acknowledge G. Carugno and E. Cornell for stimulating discussions, and I. Dorofeyev for a critical reading of the paper. We also acknowledge supports by the Ministero dell'Istruzione, dell'Universit\`a e della Ricerca (MIUR).
%  
%%%%%%%%%%%%%%%%%%%%%%%%%%%%%%%%%%%%%%%%%%%%%%%%%%%%%%%%%%%%%%%%%%%%%%%%%%%%%%%%%%%%%%%%%%%%%%%%


\begin{thebibliography}{99}
\bibitem{Spatial} The presence of the delta function is justified by the fact that the relevant distances are much larger
than the non-locality lengths (nanometers) characterizing the optical properties of the media.
\bibitem{Dorofeyevcomment} After on-line submission of the present paper I. Dorofeyev informed us that the general formalism developed in \cite{Dorofeyev1} agrees with Eqs. (\ref{ForceNeq})-(\ref{EWI}), so that we argue that results of \cite{Dorofeyev2} were based on some inconsistent derivation.
\bibitem{absorption} Neglecting the absorption actually requires  the condition
$L\ll \lambda _{T}^{2}/l\varepsilon _{2}''$. We also assume that there is no thermal radiation coming from $z=+\infty$.
\end{thebibliography}
\end{document}